# An Eshelbian approach to the nonlinear mechanics of constrained solid-fluid mixtures

**S. Quiligotti** and **G.A. Maugin**, Paris, France, and **F. dell'Isola**, Rome, Italy

**Summary.** Looking at rational mixture theories within the context of a new perspective, this work aims to put forward a proposal for an Eshelbian approach to the nonlinear mechanics of a constrained solid-fluid mixture, made up of an inhomogeneous poroelastic solid and an inviscid compressible fluid, which do not undergo any chemical reaction.

## 1 Introduction

A binary solid-fluid mixture is usually thought of as a couple of body manifolds, $\mathfrak{B}_S$ and $\mathfrak{B}_F$, embedded into the three-dimensional Euclidean space so as to occupy, in the course of their independent motions, a common smooth region of the physical environment (see, e.g., Atkin and Craine [1], Bowen [2], Rajagopal and Tao [24], Truesdell [28], [29]). If a smooth region of the space is chosen as a reference shape, which need not ever be occupied by the solid constituent, then a motion of $\mathfrak{B}_S$ can be described as a time sequence of mappings which carry the solid manifold from the reference to the current shape, whereas the motion of the fluid constituent can be conceived as a time sequence of embeddings of $\mathfrak{B}_F$ into the three-dimensional physical space.

By virtue of such a customary fundamental assumption, any place in the current shape of the mixture is simultaneously occupied by a material point belonging to each constituent. Henceforth, the motion of the fluid-body manifold may be described by taking into account that any fluid point is naturally associated with the 1-parameter family of reference places occupied by the solid points that are currently overlapped with it (see Wilmanski [31]–[33]). Accordingly, both the Eulerian fluid and solid velocity fields can be pulled back to the linear vector space associated with the reference shape of the solid constituent (Sect. 2), and a referential description of all relevant fluid properties can be furthermore introduced and motivated (Sect. 3).

In order to derive the required number of field balance equations and boundary conditions which govern the dynamics of unconstrained solid-fluid mixtures (Sect. 4), a suitable expression for the power expended by internal and external actions on any admissible test velocity field is postulated within the framework of a first-order gradient theory (Sects. 4.1–4.3), and the thorny issue of splitting the overall traction applied on the boundary of the mixture is briefly addressed (Sect. 4.3).

Recalling the duality highlighted by d'Alembert, the principle of virtual power (see, e.g. Di Carlo [11], Germain [15], [16], Maugin [19]) is finally used as a main tool to introduce the actions which expend power on the kinematical descriptors admissible in the theory [23], leading to consistent straightforward definitions of peculiar Cauchy-like [28], Piola-like [31] and Eshelby-like stress tensors. In this way we extend the notion of material action (see, e.g. Eshelby [12]–[14], Maugin [20], [21]) to any rational theory of solid-fluid mixtures.

As the investigation of the historical development of the theory of porous media seems to point out (de Boer [6], [8]), a mathematical theory of mixtures, enriched by the concept of volume fractions, also provides a suitable framework for the development of a consistent macroscopic theory of porous solids saturated with fluids (Bowen [3], [4], de Boer [7], [8], Wilmanski [33]).

If the saturation constraint is satisfied (Klisch [17], Svendsen and Hutter [26]), i.e. the volume occupied by the constituents equals the volume available to the mixture, then the stress response is determined by the motion except for an arbitrary contribution, due to the pressure reaction which arises in the material so as to maintain each constituent in contact with the other one. Within the framework of a variational theory (Sects. 4.4–4.7), the *saturation* pressure can be truly interpreted as a Lagrangian multiplier in the expression of the strain-energy density per unit volume of the mixture [9], so as to extend the concept of *effective stress* and derive, as a result of the theory, the splitting rule which governs the distribution of such a pressure among the constituents of a saturated solid-fluid mixture.

## 2 Kinematics

Let us focus our attention on the kinematics of a binary mixture $\mathfrak{B}$, consisting of two smooth three-dimensional[1] material manifolds,

$$\mathfrak{B} := \{\mathfrak{B}_S, \mathfrak{B}_F\}. \tag{1}$$

In order to avoid confusion between particles which belong to any constituent of the mixture, we refer to material points $\mathfrak{X}_\alpha \in \mathfrak{B}_\alpha$ (Fig. 1) as $\alpha$-*points*, with $\alpha \in \{S, F\}$.

By assumption [22], there exists a smooth embedding of the body manifold $\mathfrak{B}_S$ into the three-dimensional Euclidean space $\mathscr{E}$,

$$\mathscr{K}_S : \mathfrak{B}_S \to \mathscr{E}, \tag{2}$$

which associates any material $S$-point with a reference place. As the embedding $\mathscr{K}_S$ does not depend on time, a smooth motion of $\mathfrak{B}_S$ may be simply thought of as a time sequence of mappings,

$$\chi_S(\cdot, t) : \mathscr{B} \to \mathscr{E}, \tag{3}$$

which carry the body manifold $\mathfrak{B}_S$ from the reference shape $\mathscr{B} \subset \mathscr{E}$ to the current shape $\chi_S(\mathscr{B}, t) \subset \mathscr{E}$. Similarly, a smooth motion of $\mathfrak{B}_F$ may be described by a time sequence of embeddings,

$$\chi_F(\cdot, t) : \mathfrak{B}_F \to \mathscr{E}, \tag{4}$$

which map the body manifold $\mathfrak{B}_F$ onto the current shape $\chi_F(\mathfrak{B}_F, t) \subset \mathscr{E}$.

According to the classical theory of mixtures, any place $x \in \mathscr{B}_t$ in the current shape

$$\mathscr{B}_t := \{\chi_S(\mathscr{K}_S(\mathfrak{B}_S), t)\} \bigcap \{\chi_F(\mathfrak{B}_F, t)\} \tag{5}$$

---

[1] We definitely do not deal with Cantor dust (fluid drops or solid slivers) and fractals such as Menger sponges and Sierpinski gaskets.

An Eshelbian approach to constrained solid-fluid mixtures is simultaneously occupied by a material point of each constituent, $\mathfrak{X}_S \in \mathfrak{B}_S$ and $\mathfrak{X}_F \in \mathfrak{B}_F$, such that

$$x = \boldsymbol{\chi}_S(\mathscr{K}_S(\mathfrak{X}_S), t) = \boldsymbol{\chi}_F(\mathfrak{X}_F, t). \tag{6}$$

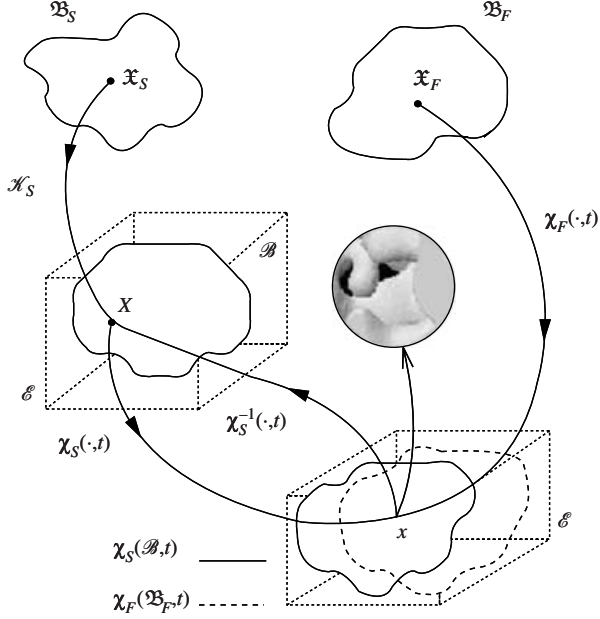

**Fig. 1.** Kinematics of a binary solid-fluid mixture

The Eulerian velocity fields

$$\mathbf{v}_\alpha(\cdot, t) : x \mapsto \mathbf{v}_\alpha(x, t), \quad \alpha \in \{S, F\} \tag{7}$$

associate the velocity pair $\mathbf{v}_S(x,t), \mathbf{v}_F(x,t)$ with the place currently occupied by $\mathfrak{X}_S$ and $\mathfrak{X}_F$. In particular, as the reference shape $\mathscr{B}$ of the material manifold $\mathfrak{B}_S$ does not depend on time, the velocity of any $S$-point can be easily obtained by taking the partial derivative of the motion $\boldsymbol{\chi}_S$ with respect to time,

$$\mathbf{v}_S(x,t) := \frac{\mathscr{D}^S}{\mathscr{D}t}[\boldsymbol{\chi}_S(\cdot, t) \circ \mathscr{K}_S](\mathfrak{X}_S) = \frac{\partial \boldsymbol{\chi}_S}{\partial t}(X, t), \tag{8}$$

where $X = \mathscr{K}_S(\mathfrak{X}_S)$ and $x = \boldsymbol{\chi}_S(X, t)$.

As we exclude a priori the possibility that a three-dimensional region of the reference shape can collapse under the motion $\boldsymbol{\chi}_S$,

$$\det \mathbf{F}_S > 0, \quad \mathbf{F}_S := \operatorname{Grad} \boldsymbol{\chi}_S = \frac{\partial \boldsymbol{\chi}_S}{\partial X}, \tag{9}$$

there exists a smooth inverse mapping,

$$\boldsymbol{\chi}_S^{-1}(\cdot, t) : \mathscr{B}_t \to \mathscr{E}, \tag{10}$$

which satisfies the trivial identity

$$X = \boldsymbol{\chi}_S^{-1}(\boldsymbol{\chi}_S(X, t), t), \quad \text{for any} \quad X \in \mathscr{B}. \tag{11}$$

Denoting by $\mathbf{V}_S$ the referential description of the partial derivative of $\boldsymbol{\chi}_S^{-1}$ with respect to time [20],

$$\mathbf{V}_S(\cdot, t) : X \mapsto \mathbf{V}_S(X, t) := \frac{\partial \boldsymbol{\chi}_S^{-1}}{\partial t}(x, t), \tag{12}$$

it can be easily shown that the following property holds:

$$\mathbf{I} = \left(\frac{\partial \boldsymbol{\chi}_S^{-1}}{\partial x} \circ \boldsymbol{\chi}_S\right) \frac{\partial \boldsymbol{\chi}_S}{\partial X} \implies \operatorname{grad} \boldsymbol{\chi}_S^{-1}\big|_{x,t} = (\operatorname{Grad} \boldsymbol{\chi}_S)^{-1}\big|_{X,t}. \tag{13}$$

Moreover, we notice that the material $S$-derivative of the identity (11) leads to the further remarkable property

$$\mathbf{F}_S \mathbf{V}_S + \frac{\partial \boldsymbol{\chi}_S}{\partial t} = \mathbf{0}, \tag{14}$$

which asserts that $\mathbf{V}_S$ is just the opposite of the pull-back of $\mathbf{v}_S$.

Bearing in mind the identity (14) and the following definition:

$$\operatorname{Grad} \mathbf{F}_S := \operatorname{Grad}(\operatorname{Grad} \boldsymbol{\chi}_S), \tag{15}$$

we can furthermore deduce that

$$\operatorname{Grad} \frac{\partial \boldsymbol{\chi}_S}{\partial t} = -(\operatorname{Grad} \mathbf{F}_S)\mathbf{V}_S - \mathbf{F}_S(\operatorname{Grad} \mathbf{V}_S). \tag{16}$$

In order to deal with a description of the motion of $F$-points through the reference shape of the solid constituent, we notice that any $F$-point which belongs to the mixture at time $t$, namely $\mathfrak{X}_F \in \boldsymbol{\chi}_F^{-1}(\mathscr{B}_t, t) \subset \mathfrak{B}_F$, interacts with a 1-parameter family of $S$-points, moving along the curve

$$\boldsymbol{\chi}_S^{-1}(\boldsymbol{\chi}_F(\mathfrak{X}_F, \cdot), \cdot) : t \mapsto X, \tag{17}$$

at the velocity $\mathbf{w}_F(X, t)$, defined by

$$\mathbf{v}_F(x, t) = \mathbf{F}_S(X, t)\, \mathbf{w}_F(X, t) + \mathbf{v}_S(x, t), \tag{18}$$

with $x = \boldsymbol{\chi}_F(\mathfrak{X}_F, t) = \boldsymbol{\chi}_S(X, t) \in \mathscr{B}_t$.

Finally, we introduce the velocity field $\mathbf{V}_F$, which identically meets the requirement

$$\mathbf{v}_F(x, t) + \mathbf{F}_S(X, t)\, \mathbf{V}_F(X, t) = \mathbf{0}, \tag{19}$$

such that

$$\mathbf{V}_F - \mathbf{V}_S = -\mathbf{w}_F. \tag{20}$$

With the aim to sketch out the role played in the theory by the velocity fields $\mathbf{V}_S$ and $\mathbf{V}_F$ [23], a few concluding remarks can be made.

Let us consider a migrating surface which envelops, at time $t$, a smooth region of the current shape of the mixture, $\mathscr{V}_t = \gamma(t) \subset \mathscr{B}_t$. If such a surface moves independently of the solid constituent, the time derivative of the integral of any smooth Eulerian scalar field $\varphi$, following the motion of the migrating surface, is given by the expression:

$$\left\{\frac{\mathrm{d}}{\mathrm{d}\tau}\int_{\gamma(\tau)} \varphi\right\}_{\tau=t} = \left\{\frac{\mathrm{d}}{\mathrm{d}\tau}\int_{\mathscr{V}_S(\tau)} \varphi\right\}_{\tau=t} + \int_{\partial \mathscr{V}_t} \varphi(\mathbf{v} - \mathbf{v}_S) \cdot \mathbf{n}, \tag{21}$$

where $\mathbf{v}$ represents the independent velocity of the moving boundary, $\mathscr{V}_S(\tau)$ the shape at time $\tau$ of the solid subbody associated with the smooth fixed region of the reference shape $\mathscr{V}^\star = \boldsymbol{\chi}_S^{-1}(\mathscr{V}_t, t)$, and $\mathbf{n}$ the outward normal to the migrating surface.

As the inverse mapping $\boldsymbol{\chi}_S^{-1}$ carries $\gamma(\tau)$ onto $\gamma^\star(\tau)$ at any time $\tau$, denoting by $\mathbf{w}$ the velocity at which the boundary $\partial \gamma^\star(\tau)$ moves through the reference shape $\mathscr{B}$,

An Eshelbian approach to constrained solid-fluid mixtures

$$\mathbf{w}(X,t) = \mathbf{F}_S^{-1}(X,t)\{\mathbf{v}(x,t) - \mathbf{v}_S(x,t)\}, \tag{22}$$

and by $\varphi^\star$ the referential description of the Eulerian scalar field $\varphi$,

$$\varphi^\star(X,t) = \det \mathbf{F}_S(X,t)\ \varphi(x,t), \tag{23}$$

we may also consider the alternative expression:

$$\left\{\frac{\mathrm{d}}{\mathrm{d}\tau}\int_{\gamma^\star(\tau)} \varphi^\star\right\}_{\tau=t} = \left\{\frac{\mathrm{d}}{\mathrm{d}\tau}\int_{\mathscr{V}_S^\star(\tau)} \varphi^\star\right\}_{\tau=t} + \int_{\partial\mathscr{V}_t^\star} \varphi^\star \mathbf{w} \cdot \mathbf{N}, \tag{24}$$

where $\mathscr{V}_S^\star(\tau) = \mathscr{V}^\star$ for any $\tau$, and $\mathscr{V}_t^\star = \gamma^\star(t) = \mathscr{V}_S^\star(t)$. If we introduce the velocity field $\mathbf{V}$, which identically meets the requirement

$$\mathbf{v}(x,t) + \mathbf{F}_S(X,t)\ \mathbf{V}(X,t) = \mathbf{0}, \tag{25}$$

such that

$$\int_{\partial\mathscr{V}_t} \varphi(\mathbf{v}-\mathbf{v}_S)\cdot\mathbf{n} = -\int_{\partial\mathscr{V}_t^\star} \varphi^\star(\mathbf{V}-\mathbf{V}_S)\cdot\mathbf{N} = \int_{\partial\mathscr{V}_t^\star} \varphi^\star \mathbf{w}\cdot\mathbf{N}, \tag{26}$$

then we can deduce [10] that, at any time $t$, the integral of the partial time derivative of the smooth scalar field $\varphi$ over $\mathscr{V}_t$ is given by the expression:

$$\int_{\mathscr{V}_t}\left\{\frac{\partial\varphi}{\partial\tau}\right\}_{\tau=t} = \left\{\frac{\mathrm{d}}{\mathrm{d}\tau}\int_{\gamma(\tau)}\varphi\right\}_{\tau=t} - \int_{\partial\mathscr{V}_t}\varphi\mathbf{v}\cdot\mathbf{n} = \left\{\frac{\mathrm{d}}{\mathrm{d}\tau}\int_{\gamma^\star(\tau)}\varphi^\star\right\}_{\tau=t} + \int_{\partial\mathscr{V}_t^\star}\varphi^\star\mathbf{V}\cdot\mathbf{N}. \tag{27}$$

Moreover, if we think of a fixed Eulerian surface $\partial\mathscr{V} \subset \mathscr{B}_t$, we notice that if the material solid surface currently overlapped with $\partial\mathscr{V}$ expands, then the migrating surface associated with $\partial\mathscr{V}$ by the inverse mapping $\chi_S^{-1}$ shrinks; conversely, if the material solid surface currently overlapped with $\partial\mathscr{V}$ shrinks, then the associated surface expands,

$$\int_{\partial\mathscr{V}} \mathbf{v}_S\cdot\mathbf{n} = -\int_{\partial\mathscr{V}_t^\star} (\det\mathbf{F}_S)\mathbf{V}_S\cdot\mathbf{N}, \quad \mathscr{V}_t^\star = \chi_S^{-1}(\mathscr{V},t). \tag{28}$$

## 3 Mass balance

Let us consider a subset of the solid material manifold, $\mathfrak{P}_S \subset \mathfrak{B}_S$, and denote by $\mathscr{V}_S^\star = \mathscr{K}_S(\mathfrak{P}_S)$ its reference shape. At any given time $t$, the solid motion $\chi_S$ carries the material subbody $\mathfrak{P}_S$ from its reference shape $\mathscr{V}_S^\star$ to its current shape, $\mathscr{V}_S(t) = \chi_S(\mathscr{V}_S^\star,t) \subset \mathscr{B}_t$.

Focusing our attention on the solid constituent, we can properly introduce at least two different mass densities per unit volume of the mixture, namely the smooth scalar fields $\varrho_S$ and $\varrho_S^\star$, defined in such a way that the solid-mass content of both $\mathscr{V}_S(t) \subset \mathscr{B}_t$ and $\mathscr{V}_S^\star \subset \mathscr{B}$ equals the measure

$$\mathscr{M}_S(\mathscr{V}_t) = \int_{\mathscr{V}_S(t)} \varrho_S = \int_{\mathscr{V}_S^\star} \varrho_S^\star, \quad \varrho_S^\star(X,t) = J_S(x,t)\varrho_S(x,t), \tag{29}$$

where $J_S(x,t) = \det\mathbf{F}_S(X,t)$ and $\mathscr{V}_t = \mathscr{V}_S(t)$.

If no phase transition between the constituents of the mixture is allowed, at any given time $\tau = t$, the value of the material $S$-derivative of $\mathscr{M}_S(\mathscr{V}_\tau)$ vanishes for any $\mathscr{V}_S(t)$,

$$\left\{\frac{d}{d\tau}\int_{\mathscr{V}_S(\tau)} \varrho_S\right\}_{\tau=t} = \int_{\mathscr{V}_S(t)}\left(\frac{\partial \varrho_S}{\partial t} + \operatorname{div}(\varrho_S \mathbf{v}_S)\right) = 0, \tag{30}$$

leading to the following expression of the local conservation law:

$$\frac{\partial \varrho_S}{\partial t} + \operatorname{div}(\varrho_S \mathbf{v}_S) = 0. \tag{31}$$

Recalling the definition of the referential mass density (29), we may also notice that the smooth scalar map $\varrho_S^\star$ does not depend on time,

$$\left\{\frac{d}{d\tau}\int_{\mathscr{V}_S(\tau)} \varrho_S\right\}_{\tau=t} = \left\{\frac{d}{d\tau}\int_{\mathscr{V}_S^\star} \varrho_S^\star\right\}_{\tau=t} = \int_{\mathscr{V}_S^\star}\left\{\frac{\partial \varrho_S^\star}{\partial \tau}\right\}_{\tau=t} = 0, \tag{32}$$

and therefore the local conservation law (31) can be rewritten in the alternative form:

$$\frac{\partial \varrho_S^\star}{\partial t} = 0. \tag{33}$$

Because of the overlapping between the two constituents, any smooth region of the current shape of the mixture can also be associated with a fluid subbody; in particular, there exists a subbody $\mathfrak{P}_F \subset \mathfrak{B}_F$ such that $\mathscr{V}_F(\tau) = \chi_F(\mathfrak{P}_F, \tau)$, with $\mathscr{V}_F(\tau) = \mathscr{V}_S(\tau)$ at time $\tau = t$. As a consequence, we can state the integral conservation law

$$\left\{\frac{d}{d\tau}\int_{\mathscr{V}_F(\tau)} \varrho_F\right\}_{\tau=t} = \int_{\mathscr{V}_F(t)}\left(\frac{\partial \varrho_F}{\partial t} + \operatorname{div}(\varrho_F \mathbf{v}_F)\right) = 0, \tag{34}$$

which yields the local expression

$$\frac{\partial \varrho_F}{\partial t} + \operatorname{div}(\varrho_F \mathbf{v}_F) = 0. \tag{35}$$

As the fluid-mass content of any smooth region of the current shape of the mixture, $\mathscr{V}_t = \mathscr{V}_F(t) \subset \mathscr{B}_t$, equals the measure [31]

$$\mathscr{M}_F(\mathscr{V}_t) = \int_{\mathscr{V}_F(t)} \varrho_F = \int_{\mathscr{V}_F^\star(t)} \varrho_F^\star, \tag{36}$$

where $\varrho_F^\star$ is the fluid-mass density per unit reference volume,

$$\varrho_F^\star(X, t) = J_S(x, t)\varrho_F(x, t), \tag{37}$$

and

$$\mathscr{V}_F^\star(t) = \chi_S^{-1}(\mathscr{V}_F(t), t) \subset \mathscr{B}, \tag{38}$$

by definition, at any given time $t$, the value of the material $F$-derivative of $\mathscr{M}_F(\mathscr{V}_\tau)$ vanishes for any $\mathscr{V}_t = \mathscr{V}_F(t)$,

An Eshelbian approach to constrained solid-fluid mixtures

$$\left\{\frac{\mathrm{d}}{\mathrm{d}\tau}\int_{\mathscr{V}_F^\star(\tau)} \varrho_F^\star\right\}_{\tau=t} = \int_{\mathscr{V}_F^\star(t)} \left(\frac{\partial \varrho_F^\star}{\partial t} + \mathrm{Div}\left(\varrho_F^\star \mathbf{w}_F\right)\right) = 0, \tag{39}$$

leading to the following expression of the local conservation law:

$$\frac{\partial \varrho_F^\star}{\partial t} + \mathrm{Div}\left(\varrho_F^\star \mathbf{w}_F\right) = 0. \tag{40}$$

Finally, let us think of both mass conservation laws (30) and (34) from a slightly different point of view, referring to the motion of the mixture as a single body [29]. In particular, let us assume that any material point of the mixture as a whole moves at the given velocity

$$\mathbf{v} := \xi_S \mathbf{v}_S + \xi_F \mathbf{v}_F, \tag{41}$$

where $\xi_\alpha$ represents the mass fraction associated with the $\alpha$-th constituent,

$$\xi_\alpha := \frac{\varrho_\alpha}{\varrho}, \quad \alpha \in \{S, F\}, \tag{42}$$

i.e. the dimensionless ratio of the current mass density of the $\alpha$-th constituent to the current mass density of the mixture per unit volume,

$$\varrho := \varrho_S + \varrho_F, \tag{43}$$

such that, by definition,

$$\xi_S + \xi_F = 1. \tag{44}$$

Taking into account the assumptions (30) and (34), we find out that the following integral conservation law of global mass holds for any smooth region of the current shape of the mixture:

$$\left\{\frac{\mathrm{d}}{\mathrm{d}\tau}\left(\int_{\mathscr{V}_S(\tau)} \varrho_S + \int_{\mathscr{V}_F(\tau)} \varrho_F\right)\right\}_{\tau=t} = \int_{\mathscr{V}_t} \left(\frac{\partial \varrho}{\partial t} + \mathrm{div}\left(\varrho \mathbf{v}\right)\right) = 0, \tag{45}$$

where, as usual, $\mathscr{V}_t = \mathscr{V}_S(t) = \mathscr{V}_F(t)$. Similarly, referring to the reference shape of the solid constituent, we can write that

$$\left\{\frac{\mathrm{d}}{\mathrm{d}\tau}\left(\int_{\mathscr{V}_S^\star} \varrho_S^\star + \int_{\mathscr{V}_F^\star(\tau)} \varrho_F^\star\right)\right\}_{\tau=t} = \int_{\mathscr{V}_t^\star} \left(\frac{\partial \varrho^\star}{\partial t} + \mathrm{Div}\left(\varrho^\star \mathbf{w}\right)\right) = 0, \tag{46}$$

where $\varrho^\star$ is the mass density of the mixture per unit reference volume,

$$\varrho^\star(X,t) := \varrho_S^\star(X,t) + \varrho_F^\star(X,t) = J_S(x,t)\varrho(x,t), \tag{47}$$

and

$$\mathbf{w} := \xi_F \mathbf{w}_F. \tag{48}$$

As a general rule, we finally notice that the time derivative of the $\alpha$-mass content of any smooth region $\mathscr{V}_\tau \subset \mathscr{B}_\tau$, enveloped by a migrating surface which follows the motion of the mixture as a whole, does not vanish at any time $\tau = t$. This means that the quantity

$$\left\{\frac{\mathrm{d}}{\mathrm{d}\tau}\int_{\mathscr{V}_\tau}\varrho_\alpha\right\}_{\tau=t} = \int_{\mathscr{V}_t}\varrho\left\{\frac{\mathscr{D}\xi_\alpha}{\mathscr{D}\tau}\right\}_{\tau=t} = -\int_{\partial\mathscr{V}_t}\varrho_\alpha(\mathbf{v}_\alpha - \mathbf{v})\cdot\mathbf{n}, \tag{49}$$

where the difference $\mathbf{v}_\alpha - \mathbf{v}$ represents the velocity of diffusion of the α-th constituent through the mixture, may be non-vanishing.

## 4 Dynamics

### 4.1 Stress power

By assumption, any given place in the current shape of the mixture is simultaneously occupied by both an $S$-point and an $F$-point (Fig. 1).

In order to describe their local interactions within the framework of a first-order gradient theory, we assume that the stress power $\omega$, expended on any pair of smooth velocity fields $(\mathbf{v}_S, \mathbf{v}_F)$, is given by the expression [16]:

$$\omega := \sum_{\alpha\in\{S,F\}}(\boldsymbol{\pi}_\alpha\cdot\mathbf{v}_\alpha + \boldsymbol{\sigma}_\alpha\cdot\operatorname{grad}\mathbf{v}_\alpha). \tag{50}$$

According to the principle of material frame-indifference, the stress power expended on any rigid-body velocity field,

$$\mathbf{v}_S(x,t) = \mathbf{v}_F(x,t) = \mathbf{w}_o(t) + \mathbf{W}(t)(x - x_o), \quad \mathbf{W}(t)\in\mathrm{Skw}, \tag{51}$$

vanishes for any choice of spatially uniform $\mathbf{w}_o(t)$ and $\mathbf{W}(t)$,

$$\mathbf{w}_o\cdot\sum_{\alpha\in\{S,F\}}\boldsymbol{\pi}_\alpha + \mathbf{W}\cdot\sum_{\alpha\in\{S,F\}}\{\boldsymbol{\sigma}_\alpha + \boldsymbol{\pi}_\alpha\otimes(x-x_o)\} = 0. \tag{52}$$

As a consequence, only the constitutive assumptions which meet the following preliminary requirements,

$$\operatorname{skw}(\boldsymbol{\sigma}_S + \boldsymbol{\sigma}_F) = \mathbf{O}, \tag{53}$$

$$\boldsymbol{\pi}_S + \boldsymbol{\pi}_F = \mathbf{0}, \tag{54}$$

can be considered admissible in the theory; in particular, while the former restriction (53) states that the sum of peculiar Cauchy-like stress tensors has to be symmetric, the latter (54) states that the force exerted on any $S$-point by the overlapped $F$-point is just the opposite of the force exerted on the $F$-point by the overlapped $F$-point.

### 4.2 Kinetic energy

Keeping in mind both definitions (43) and (41), and denoting the diffusion velocity of the α-th constituent by $\mathbf{d}_\alpha$,

An Eshelbian approach to constrained solid-fluid mixtures

$$\mathbf{d}_S := \mathbf{v}_S - \mathbf{v} = \xi_F(\mathbf{v}_S - \mathbf{v}_F), \tag{55}$$

$$\mathbf{d}_F := \mathbf{v}_F - \mathbf{v} = \xi_S(\mathbf{v}_F - \mathbf{v}_S), \tag{56}$$

we define the kinetic energy density per unit volume of the mixture, conceived as a single body in motion at the velocity $\mathbf{v}$, by the expression[2]

$$K := \frac{1}{2}\, \varrho \mathbf{v} \cdot \mathbf{v}, \tag{57}$$

which results in the sum:

$$\begin{aligned} K &= \frac{\varrho_S \varrho_F}{\varrho_S + \varrho_F}\, \mathbf{v}_S \cdot \mathbf{v}_F + \sum_{\alpha \in \{S,F\}} \frac{1}{2}\, \varrho_\alpha \mathbf{v}_\alpha \cdot \xi_\alpha \mathbf{v}_\alpha = \\ &= -\frac{1}{2}\, \frac{\varrho_S \varrho_F}{\varrho_S + \varrho_F}\, (\mathbf{v}_F - \mathbf{v}_S) \cdot (\mathbf{v}_F - \mathbf{v}_S) + \sum_{\alpha \in \{S,F\}} \frac{1}{2}\, \varrho_\alpha \mathbf{v}_\alpha \cdot \mathbf{v}_\alpha = \\ &= \frac{1}{2}\, \varrho(\mathbf{d}_S \cdot \mathbf{d}_F) + \sum_{\alpha \in \{S,F\}} \frac{1}{2}\, \varrho_\alpha \mathbf{v}_\alpha \cdot \mathbf{v}_\alpha. \end{aligned} \tag{58}$$

Accordingly, the time derivative of the kinetic energy associated with any smooth region of the current shape, enveloped by a migrating surface which follows the motion of the mixture as a single body, equals the integral of the power expended by inertial forces on the mean velocity field $\mathbf{v}$,

$$\left\{ \frac{\mathrm{d}}{\mathrm{d}\tau} \int_{\mathscr{V}_\tau} \frac{1}{2}\, \varrho \mathbf{v} \cdot \mathbf{v} \right\}_{\tau=t} = \int_{\mathscr{V}_t} \mathbf{v} \cdot \varrho \mathbf{a}, \quad \mathbf{a} := \left\{ \frac{\mathscr{D} \mathbf{v}}{\mathscr{D}\tau} \right\}_{\tau=t}, \tag{59}$$

where the differential operator $\frac{\mathscr{D}}{\mathscr{D}\tau}$ denotes the material derivative following the motion of the mixture as a single body.

It may be pointed out that, while the velocity of the centre of mass is given by the sum of peculiar velocities $\mathbf{v}_\alpha$ weighted by mass fractions $\xi_\alpha$,

$$\mathbf{v} \cdot \varrho \mathbf{a} = \sum_{\alpha \in \{S,F\}} \mathbf{v}_\alpha \cdot \varrho_\alpha \mathbf{a}, \tag{60}$$

as a general rule [28], the material derivative of $\mathbf{v}$, following the motion of the mixture as a single body, does not equal the mean of peculiar accelerations $\mathbf{a}_\alpha$, associated with the overlapped $\alpha$-points,

$$\mathbf{a} = \sum_{\alpha \in \{S,F\}} \left\{ \xi_\alpha \mathbf{a}_\alpha - \frac{1}{\varrho}\, \mathrm{div}\, (\varrho_\alpha \mathbf{d}_\alpha \otimes \mathbf{d}_\alpha) \right\}, \quad \mathbf{a}_\alpha := \left\{ \frac{\mathscr{D}^\alpha \mathbf{v}_\alpha}{\mathscr{D}\tau} \right\}_{\tau=t}, \tag{61}$$

where the differential operator $\frac{\mathscr{D}^\alpha}{\mathscr{D}\tau}$ denotes the material derivative following the $\alpha$-motion. Moreover, denoting by $\bar{\boldsymbol{\sigma}}$ the second-order tensor which takes into account the apparent stress due to diffusive motions,

$$\bar{\boldsymbol{\sigma}} := \sum_{\alpha \in \{S,F\}} \varrho_\alpha \mathbf{d}_\alpha \otimes \mathbf{d}_\alpha, \tag{62}$$

and recalling the property

---

[2] Alternative forms (and the associated drawbacks) are discussed in de Boer [8].

$$\xi_\alpha \,\mathrm{div}\,\bar{\boldsymbol{\sigma}} = \mathrm{div}\,(\xi_\alpha \bar{\boldsymbol{\sigma}}) - \bar{\boldsymbol{\sigma}}\,\mathrm{grad}\,\xi_\alpha, \quad \forall \alpha \in \{S,F\}, \tag{63}$$

the expression (61) finally results in:

$$\varrho_S \mathbf{a} = \varrho_S \mathbf{a}_S - \mathrm{div}\,(\xi_S \bar{\boldsymbol{\sigma}}) + \frac{\varrho_S \varrho_F}{\varrho_S + \varrho_F}(\mathbf{a}_F - \mathbf{a}_S) + \bar{\boldsymbol{\sigma}}\,\mathrm{grad}\,\xi_S, \tag{64}$$

$$\varrho_F \mathbf{a} = \varrho_F \mathbf{a}_F - \mathrm{div}\,(\xi_F \bar{\boldsymbol{\sigma}}) + \frac{\varrho_S \varrho_F}{\varrho_S + \varrho_F}(\mathbf{a}_S - \mathbf{a}_F) + \bar{\boldsymbol{\sigma}}\,\mathrm{grad}\,\xi_F. \tag{65}$$

### 4.3 The principle of virtual power

So as to deduce the required number of local balance equations which govern the dynamics of the mixture, we state that the total power expended vanishes on any conceivable smooth *test* velocity field $\hat{\mathbf{v}}_\alpha$,

$$\sum_{\alpha \in \{S,F\}} \left\{ \int_{\mathcal{B}_t} \varrho_\alpha(\mathbf{f} - \mathbf{a}) \cdot \hat{\mathbf{v}}_\alpha + \int_{\partial \mathcal{B}_t} \xi_\alpha \mathbf{t} \cdot \hat{\mathbf{v}}_\alpha \right\} - \sum_{\alpha \in \{S,F\}} \left\{ \int_{\mathcal{B}_t} (\boldsymbol{\pi}_\alpha \cdot \hat{\mathbf{v}}_\alpha + \boldsymbol{\sigma}_\alpha \cdot \mathrm{grad}\,\hat{\mathbf{v}}_\alpha) \right\} = 0, \tag{66}$$

where $\mathbf{f}$ is the applied external force per unit mass of the mixture,

$$\varrho \mathbf{f} = \sum_{\alpha \in \{S,F\}} \varrho_\alpha \mathbf{f}, \tag{67}$$

and $\mathbf{t}$ the overall external boundary traction.

Keeping in mind that, for any $\alpha \in \{S, F\}$,

$$\boldsymbol{\sigma}_\alpha \cdot \mathrm{grad}\,\hat{\mathbf{v}}_\alpha = \mathrm{div}\,(\boldsymbol{\sigma}_\alpha^T \hat{\mathbf{v}}_\alpha) - \hat{\mathbf{v}}_\alpha \cdot \mathrm{div}\,\boldsymbol{\sigma}_\alpha, \tag{68}$$

the integral equation (66) leads to the set of local equations:

$$\mathrm{div}\,\boldsymbol{\sigma}_\alpha + \varrho_\alpha \mathbf{f} - \boldsymbol{\pi}_\alpha = \varrho_\alpha \mathbf{a}, \tag{69}$$

$$\boldsymbol{\sigma}_\alpha \mathbf{n} = \xi_\alpha \mathbf{t}. \tag{70}$$

Consistently [28], taking the sum over $\alpha$ and recalling the requirements (53) and (54), it is possible to show that the dynamics of the mixture as a single body results to be governed by the equations

$$\mathrm{div}\,\boldsymbol{\sigma} + \varrho \mathbf{f} = \varrho \mathbf{a}, \tag{71}$$

$$\boldsymbol{\sigma} \mathbf{n} = \mathbf{t}, \tag{72}$$

with

$$\boldsymbol{\sigma} := \sum_{\alpha \in \{S,F\}} \boldsymbol{\sigma}_\alpha, \quad \boldsymbol{\sigma} \in \mathrm{Sym}. \tag{73}$$

As the tricky question of splitting the overall applied boundary traction $\mathbf{t}$ (72) among the constituents still stands as one of the greatest challenges that have to be faced up in order to put mixture theories to use [24], it is worth recalling that, within the framework of variational principles, boundary conditions are straightforwardly derived, as well as governing equations, as a result of the theory. All conclusions drawn here might be consistently extended to higher-order gradient theories, provided that a meaningful physical interpretation of further emerging boundary conditions can be taken for granted [9]. Nevertheless, such an approach is far from general (see, e.g. Reynolds and Humphrey [25]), as well as other available approaches [5], [24], [33] do not exhaust the list of possibilities that need to be considered, for instance, within the framework of biomechanical applications of mixture theories [27].

An Eshelbian approach to constrained solid-fluid mixtures

Finally, bearing in mind the expressions (64) and (65), and denoting by $\tilde{\boldsymbol{\sigma}}_\alpha$ and $\tilde{\mathbf{f}}_\alpha$, respectively, the $\alpha$-th peculiar stress tensor [2], [29], [30] and the external force per unit mass of the $\alpha$-th constituent,

$$\tilde{\boldsymbol{\sigma}}_\alpha := \boldsymbol{\sigma}_\alpha + \xi_\alpha \bar{\boldsymbol{\sigma}}, \tag{74}$$

$$\varrho_\alpha \tilde{\mathbf{f}}_\alpha := \varrho_\alpha \mathbf{f} + \varrho_\alpha (\mathbf{a}_\alpha - \mathbf{a}) - \boldsymbol{\pi}_\alpha - \operatorname{div}(\xi_\alpha \bar{\boldsymbol{\sigma}}), \tag{75}$$

Cauchy's first law of motion (69) and the expression of the symmetric stress tensor (73) respectively, result in

$$\operatorname{div} \tilde{\boldsymbol{\sigma}}_\alpha + \varrho_\alpha \tilde{\mathbf{f}}_\alpha = \varrho_\alpha \mathbf{a}_\alpha \tag{76}$$

and

$$\boldsymbol{\sigma} = \sum_{\alpha \in \{S, F\}} (\tilde{\boldsymbol{\sigma}}_\alpha - \xi_\alpha \bar{\boldsymbol{\sigma}}) = \sum_{\alpha \in \{S, F\}} (\tilde{\boldsymbol{\sigma}}_\alpha - \varrho_\alpha \mathbf{d}_\alpha \otimes \mathbf{d}_\alpha). \tag{77}$$

### 4.4 Constitutive prescriptions: a variational approach

In order to investigate the configurational nature of hyperelastic interactions, let us assume that there exists a strain-energy density per unit volume of the current shape of the mixture, such that [9] the material time derivative of the strain-energy content, associated with any fit region which follows the motion of the mixture,

$$\left\{ \frac{\mathrm{d}}{\mathrm{d}\tau} \int_{\mathscr{V}_\tau} W \right\}_{\tau = t} = \int_{\mathscr{V}_t} \left( \frac{\partial W}{\partial t} + \operatorname{div}(W\mathbf{v}) \right), \tag{78}$$

is equal to the stress power expended on the velocity pair $(\mathbf{v}_S, \mathbf{v}_F)$,

$$\left\{ \frac{\mathrm{d}}{\mathrm{d}\tau} \int_{\mathscr{V}_\tau} W \right\}_{\tau = t} = \sum_{\alpha \in \{S, F\}} \int_{\mathscr{V}_t} (\boldsymbol{\pi}_\alpha \cdot \mathbf{v}_\alpha + \boldsymbol{\sigma}_\alpha \cdot \operatorname{grad} \mathbf{v}_\alpha). \tag{79}$$

As the previous integral definition holds for any choice of $\mathscr{V}_t \subset \mathscr{B}_t$, we can properly localize it on the current shape of the mixture,

$$\frac{\partial W}{\partial t} + W \operatorname{div} \mathbf{v} + \mathbf{v} \cdot \operatorname{grad} W = \sum_{\alpha \in \{S, F\}} (\boldsymbol{\pi}_\alpha \cdot \mathbf{v}_\alpha + \boldsymbol{\sigma}_\alpha \cdot \operatorname{grad} \mathbf{v}_\alpha). \tag{80}$$

### 4.5 The configurational nature of interactions

Let us introduce a partial Piola-Kirchhoff stress tensor $\mathbf{T}_\alpha$ for each constituent of the mixture [31], such that, for any $\mathscr{V} \subset \mathscr{B}$,

$$\int_{\mathscr{V}} \mathbf{T}_\alpha \cdot \operatorname{Grad} \mathbf{v}_\alpha = \int_{\mathscr{V}_t} \boldsymbol{\sigma}_\alpha \cdot \operatorname{grad} \mathbf{v}_\alpha, \quad \mathscr{V}_t = \boldsymbol{\chi}_S(\mathscr{V}, t), \tag{81}$$

i.e.

$$\mathbf{T}_\alpha = J_S \boldsymbol{\sigma}_\alpha \mathbf{F}_S^{-T}, \quad \alpha \in \{S, F\}. \tag{82}$$

By assumption, the time derivative of the stored energy associated with any fit region of the reference shape, enveloped by a migrating surface which follows the motion of the mixture as a whole, equals the power expended by internal actions,

$$\left\{ \frac{\mathrm{d}}{\mathrm{d}\tau} \int_{\mathscr{V}_\tau^\star} J_S W \right\}_{\tau=t} = \sum_{\alpha \in \{S,F\}} \int_{\mathscr{V}_t^\star} (J_S \boldsymbol{\pi}_\alpha \cdot \mathbf{v}_\alpha + \mathbf{T}_\alpha \cdot \mathrm{Grad}\, \mathbf{v}_\alpha). \tag{83}$$

Bearing in mind the identity (27), let us consider the possibility to define a further set of dynamical descriptors, $\mathbf{b}_\alpha$ and $\boldsymbol{\tau}_\alpha$, which satisfy the identity

$$\sum_{\alpha \in \{S,F\}} \int_{\mathscr{B}} (J_S \boldsymbol{\pi}_\alpha \cdot \hat{\mathbf{v}}_\alpha + \mathbf{T}_\alpha \cdot \mathrm{Grad}\, \hat{\mathbf{v}}_\alpha) + \int_{\partial \mathscr{B}} J_S W \, \hat{\mathbf{V}} \cdot \mathbf{N} =$$
$$= \sum_{\alpha \in \{S,F\}} \int_{\mathscr{B}} \left( \boldsymbol{\tau}_\alpha \cdot \hat{\mathbf{V}}_\alpha + \mathbf{b}_\alpha \cdot \mathrm{Grad}\, \hat{\mathbf{V}}_\alpha \right), \tag{84}$$

for any choice of smooth velocity fields $\hat{\mathbf{V}}_\alpha$ and $\hat{\mathbf{v}}_\alpha$, such that $\hat{\mathbf{v}}_\alpha + \mathbf{F}_S \hat{\mathbf{V}}_\alpha = \mathbf{0}$ for any $\alpha \in \{S, F\}$, and $\hat{\mathbf{V}} = \xi_S \hat{\mathbf{V}}_S + \xi_F \hat{\mathbf{V}}_F$. As a consequence, we obtain that

$$\mathbf{b}_\alpha = \xi_\alpha J_S W \, \mathbf{I} - \mathbf{F}_S^T \mathbf{T}_\alpha, \tag{85}$$

$$\boldsymbol{\tau}_\alpha = \mathrm{Grad}(\xi_\alpha J_S W) - (\mathrm{Grad}\, \mathbf{F}_S)^T \mathbf{T}_\alpha - J_S \mathbf{F}_S^T \boldsymbol{\pi}_\alpha, \tag{86}$$

where $\mathbf{b}_\alpha$ represents the *peculiar Eshelby stress tensor* (see, e.g. Eshelby [12]–[14], Maugin [20], [21]), associated with the $\alpha$-th constituent of the mixture. Moreover, in order to meet the requirement (53), it may be pointed out that the sum of peculiar stress tensors (85) needs to be symmetric with respect to the right Cauchy-Green tensor $\mathbf{C}_S = \mathbf{F}_S^T \mathbf{F}_S$ (Quiligotti [23], cf. Maugin [19]), i.e.,

$$\mathrm{skw}\,[(\mathbf{b}_S + \mathbf{b}_F)\mathbf{C}_S] = \mathbf{O}. \tag{87}$$

4.6 Unconstrained solid-fluid mixtures

So as to describe the macroscopic interactions exchanged by an inhomogeneous poroelastic solid and a compressible inviscid fluid (Krishnaswamy and Batra [18], Rajagopal and Tao [24], Svensen and Hutter [26]), we assume that the value of the strain-energy density, $W(x,t)$, depends on the value of both the gradient of the $S$-motion, referred to the $S$-point which occupies the current place $x$ at time $t$, and the fluid-mass density per unit volume of the mixture, referred to the overlapped $F$-point. Finally, as the solid body is assumed to be inhomogeneous, a further dependence of the strain-energy density on reference places is taken into account, i.e.,

$$W(\mathbf{F}_S(X,t), \varrho_F(x,t), X), \quad X = \boldsymbol{\chi}_S^{-1}(x,t). \tag{88}$$

This hypothesis yields the following constitutive prescriptions for both $\pi$-like interactions:

$$\boldsymbol{\pi}_S^{(u)} = (\xi_S - 1)(\mathrm{grad}\, \mathbf{F}_S)^T \frac{\partial W}{\partial \mathbf{F}_S} + W \,\mathrm{grad}\, \xi_S + \xi_S \frac{\partial W}{\partial \varrho_F} \,\mathrm{grad}\, \varrho_F + (\xi_S - 1)\mathbf{F}_S^{-T} \frac{\partial W}{\partial X}, \tag{89}$$

$$\boldsymbol{\pi}_F^{(u)} = \xi_F (\mathrm{grad}\, \mathbf{F}_S)^T \frac{\partial W}{\partial \mathbf{F}_S} + W \,\mathrm{grad}\, \xi_F + (\xi_F - 1) \frac{\partial W}{\partial \varrho_F} \,\mathrm{grad}\, \varrho_F + \xi_F \mathbf{F}_S^{-T} \frac{\partial W}{\partial X}, \tag{90}$$

and the $\sigma$-like interactions

An Eshelbian approach to constrained solid-fluid mixtures

$$\boldsymbol{\sigma}_F^{(u)} = -\left(\varrho_F \frac{\partial W}{\partial \varrho_F} - \xi_F W\right) \mathbf{I}, \tag{91}$$

$$\boldsymbol{\sigma}_S^{(u)} = \frac{\partial W}{\partial \mathbf{F}_S} \mathbf{F}_S^T + \xi_S W \mathbf{I}. \tag{92}$$

It may be pointed out that, while the sum of $\pi$-like interactions satisfies the condition (54), the requirement (53) should be met by taking into account that the stress tensor $\boldsymbol{\sigma}$ results in

$$\boldsymbol{\sigma}_S^{(u)} + \boldsymbol{\sigma}_F^{(u)} = \frac{\partial W}{\partial \mathbf{F}_S} \mathbf{F}_S^T + \varrho_F \left(\frac{W}{\varrho_F} - \frac{\partial W}{\partial \varrho_F}\right) \mathbf{I}. \tag{93}$$

Accordingly, the identity (85) leads to the following straightforward expressions for the peculiar Eshelby stress tensors:

$$\mathbf{b}_S^{(u)} = -J_S \mathbf{F}_S^T \frac{\partial W}{\partial \mathbf{F}_S}, \quad \text{and} \quad \mathbf{b}_F^{(u)} = J_S \varrho_F \frac{\partial W}{\partial \varrho_F} \mathbf{I}, \tag{94}$$

whose definitions seem to corroborate the importance of the role played by partial chemical potentials within the context of solid-fluid mixture theories (see Bowen [2]–[4]). Finally, the identity (86) yields the following expressions of $\tau$-like interactions:

$$\boldsymbol{\tau}_S^{(u)} = J_S \frac{\partial W}{\partial X}, \quad \text{and} \quad \boldsymbol{\tau}_F^{(u)} = \frac{\partial W}{\partial \varrho_F} \, \mathrm{Grad} \, (J_S \varrho_F), \tag{95}$$

for which the contributions given by partial derivatives of the strain-energy density with respect to the state variables $\varrho_F$ and $\mathbf{F}_S$ are uncoupled.

### 4.7 The saturation constraint

So as to develop a consistent macroscopic theory of saturated poroelastic media (see, e.g. Bowen [3], [4], Svendsen and Hutter [26], Klisch [17]), let us enrich the mathematical theory of binary solid-fluid mixtures with the introduction of the concept of volume fractions,

$$v_\alpha := \frac{\varrho_\alpha}{\hat{\varrho}_\alpha},$$

given by the dimensionless ratio of the *macroscopic* mass density $\varrho_\alpha$ to the *microscopic* mass density $\hat{\varrho}_\alpha$, which depends on the usual state variables (Fillunger, see de Boer [6], dell'Isola et al. [9]),

$$\hat{\varrho}_\alpha(\mathbf{F}_S(X,t), \varrho_F(x,t), X), \quad X = \boldsymbol{\chi}_S^{-1}(x,t). \tag{96}$$

A poroelastic solid infused with a compressible fluid is *saturated* if the solid skeleton is perfectly permeated by the fluid, i.e. if the *saturation constraint* is satisfied,

$$v_S + v_F - 1 = 0. \tag{97}$$

As constraints are naturally associated with reactive actions, a *saturation pressure* $p$, which does not expend power on any motion compatible with the constraint (97), arises in the material so as to maintain each constituent in contact with the other one. We think of such a pressure as a Lagrangian multiplier in the expression of the strain-energy density [9],

$$W + p(v_S + v_F - 1), \tag{98}$$

and generalize the effective stress principle, bearing in mind that

$$\frac{\partial v_S}{\partial \mathbf{F}_S} = -v_S \left(\mathbf{F}_S^{-T} + \frac{v_S}{\varrho_S} \frac{\partial \hat{\varrho}_S}{\partial \mathbf{F}_S}\right), \tag{99}$$

$$\frac{\partial v_F}{\partial \mathbf{F}_S} = -\frac{v_F^2}{\varrho_F} \frac{\partial \hat{\varrho}_F}{\partial \mathbf{F}_S}, \tag{100}$$

$$\frac{\partial v_S}{\partial \varrho_F} = -\frac{v_S^2}{\varrho_S} \frac{\partial \hat{\varrho}_S}{\partial \varrho_F}, \tag{101}$$

and

$$\frac{\partial v_F}{\partial \varrho_F} = \frac{v_F}{\varrho_F}\left(1 - v_F \frac{\partial \hat{\varrho}_F}{\partial \varrho_F}\right). \tag{102}$$

Neglecting the dependence on $X$ in (88) and (96) for the sake of simplicity, we find out the following expressions of $\pi$-like interactions:

$$\boldsymbol{\pi}_S^{(c)} = \hat{\boldsymbol{\pi}}_S^{(u)} + pv_F \frac{\xi_S}{\varrho_F}\left(1 - v_F \frac{\partial \hat{\varrho}_F}{\partial \varrho_F} - \frac{v_S^2}{v_F}\frac{\varrho_F}{\varrho_S}\frac{\partial \hat{\varrho}_S}{\partial \varrho_F}\right) \operatorname{grad} \varrho_F$$

$$- pv_S(\xi_S - 1)(\operatorname{grad} \mathbf{F}_S)^T\left(\mathbf{F}_S^{-T} + \frac{v_S}{\varrho_S}\frac{\partial \hat{\varrho}_S}{\partial \mathbf{F}_S} + \frac{v_F^2}{v_S}\frac{1}{\varrho_F}\frac{\partial \hat{\varrho}_F}{\partial \mathbf{F}_S}\right), \tag{103}$$

$$\boldsymbol{\pi}_F^{(c)} = \hat{\boldsymbol{\pi}}_F^{(u)} + p(\xi_F - 1)\frac{v_F}{\varrho_F}\left(1 - v_F \frac{\partial \hat{\varrho}_F}{\partial \varrho_F} - \frac{v_S^2}{v_F}\frac{\varrho_F}{\varrho_S}\frac{\partial \hat{\varrho}_S}{\partial \varrho_F}\right) \operatorname{grad} \varrho_F$$

$$- pv_S\xi_F(\operatorname{grad} \mathbf{F}_S)^T\left(\mathbf{F}_S^{-T} + \frac{v_S}{\varrho_S}\frac{\partial \hat{\varrho}_S}{\partial \mathbf{F}_S} + \frac{v_F^2}{v_S}\frac{1}{\varrho_F}\frac{\partial \hat{\varrho}_F}{\partial \mathbf{F}_S}\right), \tag{104}$$

where $\hat{\boldsymbol{\pi}}_S^{(u)}$ and $\hat{\boldsymbol{\pi}}_F^{(u)}$ are given by the general relations (89) and (90), assuming that the partial derivative of $W$ with respect to $X$ vanishes, i.e., that the value of the stored energy density (88) does not explicitly depend on $X$.

Furthermore, the $\sigma$-like interactions result in the following expressions:

$$\boldsymbol{\sigma}_F^{(c)} = \hat{\boldsymbol{\sigma}}_F^{(u)} - pv_F\left(1 - v_F \frac{\partial \hat{\varrho}_F}{\partial \varrho_F} - \frac{v_S^2}{v_F}\frac{\varrho_F}{\varrho_S}\frac{\partial \hat{\varrho}_S}{\partial \varrho_F}\right)\mathbf{I}, \tag{105}$$

$$\boldsymbol{\sigma}_S^{(c)} = \hat{\boldsymbol{\sigma}}_S^{(u)} - pv_S\left(\mathbf{F}_S^{-T} + \frac{v_S}{\varrho_S}\frac{\partial \hat{\varrho}_S}{\partial \mathbf{F}_S} + \frac{v_F^2}{v_S}\frac{1}{\varrho_F}\frac{\partial \hat{\varrho}_F}{\partial \mathbf{F}_S}\right)\mathbf{F}_S^T, \tag{106}$$

where $\hat{\boldsymbol{\sigma}}_S^{(u)}$ and $\hat{\boldsymbol{\sigma}}_F^{(u)}$ are derived, respectively, by the relations (92) and (91).

As a general result, it may be pointed out that the saturation pressure is distributed among the constituents proportionally to their volume fractions only if they are microscopically incompressible [9], i.e., if the value of microscopic mass densities $\hat{\varrho}_\alpha(x, t)$ is independent of the value of both the macroscopic fluid-mass density ($\varrho_F(x, t)$) and the gradient of the $S$-motion ($\mathbf{F}_S(X, t)$).

Finally, we get the expressions of partial Eshelby stress tensors,

$$\mathbf{b}_F^{(c)} = \hat{\mathbf{b}}_F^{(u)} + pv_F J_S\left(1 - v_F \frac{\partial \hat{\varrho}_F}{\partial \varrho_F} - \frac{v_S^2}{v_F}\frac{\varrho_F}{\varrho_S}\frac{\partial \hat{\varrho}_S}{\partial \varrho_F}\right)\mathbf{I}, \tag{107}$$

$$\mathbf{b}_S^{(c)} = \hat{\mathbf{b}}_S^{(u)} + pv_S J_S \mathbf{F}_S^T\left(\mathbf{F}_S^{-T} + \frac{v_S}{\varrho_S}\frac{\partial \hat{\varrho}_S}{\partial \mathbf{F}_S} + \frac{v_F^2}{v_S}\frac{1}{\varrho_F}\frac{\partial \hat{\varrho}_F}{\partial \mathbf{F}_S}\right), \tag{108}$$

and $\tau$-like interactions,

$$\boldsymbol{\tau}_S^{(c)} = \hat{\boldsymbol{\tau}}_S^{(u)} = \mathbf{0}, \tag{109}$$

An Eshelbian approach to constrained solid-fluid mixtures

$$\boldsymbol{\tau}_F^{(c)} = \hat{\boldsymbol{\tau}}_F^{(u)} + p \frac{v_F}{\varrho_F} \left( 1 - v_F \frac{\partial \hat{\varrho}_F}{\partial \varrho_F} - \frac{v_S^2}{v_F} \frac{\varrho_F}{\varrho_S} \frac{\partial \hat{\varrho}_S}{\partial \varrho_F} \right) \text{Grad } (J_S \varrho_F), \tag{110}$$

where $\hat{\mathbf{b}}_\alpha^{(u)}$ and $\hat{\boldsymbol{\tau}}_\alpha^{(u)}$ are derived, respectively, by the relations (94) and (95).

## 5 Conclusions

In order to put forward a proposal for an Eshelbian approach to a first-order gradient theory of constrained solid-fluid mixtures, the Eulerian velocity fields associated with each of the constituents have been pulled back to the reference shape of the solid one.

Using the principle of virtual power as a main tool, further dynamical descriptors have been introduced, by duality, and the role that they play in the theory has been sketched out within the framework of a variational approach to the dynamics of both unconstrained inhomogeneous and saturation-constrained homogeneous poroelastic solids, infused with compressible inviscid fluids.

## Acknowledgements

This work has been developed within the framework of the TMR European Network on "Phase Transitions in Crystalline Solids". The anonymous referee's advises are gratefully acknowledged.

**Authors' addresses:** S. Quiligotti and G. A. Maugin, Laboratoire de Modélisation en Mécanique, Université Pierre et Marie Curie, Case 162, 4 place Jussieu, 75252 Paris, France (E-mail: quiligotti@lmm.jussieu.fr); F. dell'Isola, Dipartimento di Ingegneria Strutturale e Geotecnica, Università di Roma "La Sapienza", Via Eudossiana 18, 00184 Roma, Italy